\documentclass[12pt]{article}

\usepackage{epsfig,multicol,multirow,cite}
\usepackage{amsmath,subfigure,latexsym,amssymb}

\def\lsi{\raise0.3ex\hbox{$<$\kern-0.75em\raise-1.1ex\hbox{$\sim$}}}
\def\gsi{\raise0.3ex\hbox{$>$\kern-0.75em\raise-1.1ex\hbox{$\sim$}}}
\def\backder{\raise1.4ex\hbox{$\leftarrow$\kern-0.75em\raise-1.4ex\hbox{$\partial$}}}
\def\forder{\raise1.4ex\hbox{$\rightarrow$\kern-0.75em\raise-1.4ex\hbox{$\partial$}}}

\newcommand{\gsim}{\mathop{\gsi}}

\newcommand{\be}{\begin{equation}}
\newcommand{\ee}{\end{equation}}
\newcommand{\nn}{\nonumber}
\newcommand{\bea}{\begin{eqnarray}}
\newcommand{\eea}{\end{eqnarray}}

\newcommand{\Z}{Z \!\!\! Z}
\newcommand{\R}{{\kern+.25em\sf{R}\kern-.78em\sf{I} \kern+.78em\kern-.25em}}
\newcommand{\N}{{\kern+.25em\sf{N}\kern-.78em\sf{I} \kern+.78em\kern-.25em}}
\newcommand{\C}{{\kern+.25em\sf{C}\kern-.50em\sf{I} \kern+.50em\kern-.25em}}

\makeatletter
\@addtoreset{equation}{section}
\makeatother

\begin{document}

{
\begin{flushright}
{\small HU-EP-06/47} \\ {\small DESY-06-234}
\end{flushright}
}

\vspace{0mm}

\begin{center}

{\Large{\bf Spin chain simulations}} \\

\vspace*{5mm}

{\Large\bf with a meron cluster algorithm} 

\vspace*{1cm}

Thomas Boyer$^{\rm \, a,b}$, \
Wolfgang Bietenholz$^{\rm \, a,c}$ \  
and \ Jair Wuilloud$^{\rm \, a,d}$ \footnote{Present address:
Westf\"{a}lische Wilhems Universit\"{a}t M\"{u}nster, \\
Inst.\ f\"{u}r Theor.\ Physik I, Wilhelm-Klemm-Str.\ 9,
D-48149 M\"{u}nster, Germany}\\

\vspace*{8mm}
{\small 
$^{\rm a}$ Institut f\"{u}r Physik,
Humboldt-Universit\"{a}t zu Berlin \\
Newtonstr.\ 15, D-12489 Berlin, Germany \\

\vspace*{4mm}

$^{\rm b}$ \'{E}cole Normale Sup\'{e}rieure de Cachan \\
61, avenue du Pr\'{e}sident Wilson \\
F-94235 Cachan Cedex, France \\

\vspace*{4mm}
$^{\rm c}$ NIC / DESY Zeuthen \\
Platanenallee 6, D-15738 Zeuthen, Germany \\

\vspace*{4mm}

$^{\rm d}$
D\'{e}partement de Physique Th\'{e}orique, 
Universit\'{e} de Gen\`{e}ve \\
24, Quai Ernest-Ansermet, CH-1211 Gen\`{e}ve 4, Switzerland \\
}

\end{center}

\vspace*{0.5cm}

We apply a meron cluster algorithm
to the XY spin chain, which describes a quantum rotor.
This is a multi-cluster simulation supplemented by 
an improved estimator, which deals with objects of 
half-integer topological charge. 
This method is powerful enough to provide precise results for the
model with a $\theta$-term --- it is therefore one of the
rare examples, where a system with a complex action
can be solved numerically. In particular we measure
the correlation length, as well as the topological and magnetic 
susceptibility. We discuss the algorithmic efficiency
in view of the critical slowing down. 
Due to the excellent performance that we observe, it is
strongly motivated to work on new applications of 
meron cluster algorithms in higher dimensions.

\newpage

\section{Introduction}

The functional integral formalism of quantum physics deals with 
infinite dimensional integrals, which can only be computed
explicitly in a few simple situations. A non-perturbative method to 
tackle models, which are not analytically soluble,
starts by a regularisation to a finite number of degrees of freedom.
This is usually achieved by a lattice discretisation of the
time (in quantum mechanics) or of the space-time (in quantum
field theory). In a finite volume the functional integral is then
given by a finite set of single variable integrals. One tries to
compute them numerically and --- based on a variety of such results ---
to extrapolate to the continuum and to infinite volume.
The transition to Euclidean time is very helpful to speed up the
convergence of the integrals.

However, the number of integrals still tends to be so large that
straight numerical integration is hopeless. Instead one
performs Monte Carlo simulations to generate a set of paths
or field configurations with the Boltzmann probability distribution
given by the Euclidean action.
Thus one evaluates the expectation values of the observables of 
interest directly at finite interaction strength
(in contrast to perturbation theory). 
On the other hand, one has to face errors due to the limited
statistics and uncertainties in the extrapolations.

Hence it is essential to optimise the algorithmic tools for such
simulations. The Metropolis algorithm is the most established
procedure, but in many cases it is far from optimal.
It can be refined to {\em cluster algorithms} \cite{SW,Wolff} which are 
by far more efficient for some set of models
(for a review, see Ref.\ \cite{Ferenc}). Unfortunately this set, 
where it could be applied successfully, is still quite small ---
in particular it excludes gauge theories up to now.\footnote{There
have been proposals for cluster algorithms for $U(1)$ gauge theory
\cite{U1}, but a breakthrough in the performance is still outstanding.
For the treatment of a discrete gauge group, see {\it e.g.}\ 
Ref.\ \cite{LQLR}.}
But in the light of the striking success in specific spin models, it is 
highly motivated to explore cluster algorithms further. 
Here we present successful tests on (classical) spin 
chains, which describe quantum mechanical systems.\footnote{The 
motivation we are giving here are functional
integrals in quantum physics, but 
cluster algorithms have a much broader range
of applicability, which also reaches out to fields like solid state
physics and biology; for recent examples, 
see Refs.\ \cite{ToOk,Mexi,DGB,LQLR}.}

Unlike the Metropolis algorithm, cluster algorithms do not
proceed from one configuration to the next by updating single spins, 
but by flipping whole clusters of them. 
First, this is promising in view of the thermalisation
time needed in the beginning of a simulation.
Later one expects to generate with the cluster algorithm
well de-correlated configurations (which are needed
for the measurements) with a modest number of simulation steps. 
In addition, multi-cluster algorithms can often be combined with
an ``improved estimator'', which allows for the inclusion of lots
of configurations that do not need to be Monte Carlo
generated explicitly.
All these properties help to reduce the computer time required to measure
an observable numerically to a given accuracy. This will be clearly 
confirmed for the system under consideration in this work, in particular 
as one approaches the continuum limit.

We apply this technique to the $O(2)$ spin chain
(the XY model), to be described in Section 2. There is a
neat way to attach a half-integer topologi\-cal charge to each cluster. 
This is the basis of a {\em meron cluster 
simulation,} which was first applied to a 2d $O(3)$ model
on a triangular lattice (with a constrained maximal angle between
neighbouring spins) \cite{merons}, see also Ref.\ \cite{brecht}.
In this framework an improved estimator is extremely powerful. 
Variants of the meron cluster algorithm can also handle
fermionic spin models successfully \cite{fermi}.

In Sections 3 and 4 we present a novel application of this algorithm
\cite{Thomas}. It enables us to
approach the continuum limit much better than the Metropolis 
algorithm, and to suppress the notorious ``critical slowing down''.
As in the original application, it is powerful enough to even explore
the system with a $\theta$-term. In almost all cases the simulation of
a system with a complex Euclidean action is hardly feasible so far
(perhaps up to a region of very small imaginary parts).\footnote{For
reviews of the situation in QCD at finite baryon density,
we refer to Refs.\ \cite{QCDnB}.}
Here we present one of the rare exceptions. Moreover, the constraint
on the angles is not necessary in our case; the latter was 
required for technical reasons in the original application \cite{merons}
(though it did not affect the universality class).
Section 5 is dedicated to our conclusions and an outlook on potential
applications of the techniques discussed here to a system of light
quarks at high temperature.

\section{The quantum rotor}

We deal with a free scalar particle of mass $m$
on a circle of radius~$1$, {\it i.e.}\ a {\em quantum rotor.} 
Its position is given by an angle $\varphi (t)$,
where $t$ is the Euclidean time, so the Lagrangian reads
$ {\bf L} = \frac{m}{2} \dot \varphi (t)^{2}$. 
We consider the propagator $G$ between the end-points 
$\varphi (0) =0 $ and $\varphi (T) = 0$ $(T>0)$, {\it i.e.}\
we assume periodic boundary conditions.
In the path integral formulation it can be decomposed
into disjoint contributions with different winding numbers $\nu$.
It is therefore the simplest quantum system with topological sectors.
As in QCD we can also insert a $\theta$-term in this summation, which
leads to
\bea
G (0,T; 0,0) &=& \sum_{\nu = -\infty}^{+\infty}
G_{\nu} (0,T; 0,0) \, e^{i \nu \theta} \nn \\
&=& \sqrt{\frac{m}{2 \pi T}} \sum_{\nu = -\infty}^{\infty}
\exp \Big[ - \frac{2m\pi^{2}}{T} \nu^{2}
+ i \nu \theta \Big] \nn \\
&=& \frac{1}{2\pi} \sum_{\nu = -\infty}^{\infty} \exp \Big[
- \frac{T}{2m} \Big( \nu - \frac{\theta}{2\pi} \Big)^{2} \Big] \ .
\eea
$G_{\nu}$ is the free propagator 
$\langle 2\pi \nu, T | 0, 0 \rangle$ on a line, 
and we set $\hbar =1$. 
It is sufficient to consider $\theta \in [0 , \pi ]$.

In QCD it appears natural that a $\theta$-term should occur,
hence it is a mystery 
--- known as the ``strong CP problem'' ---
why the observed $\theta$-angle is zero (or very close to it).
In our case, a finite $\theta$-angle
does describe a physical situation, if we assume the particle to
carry an electric charge $q$ and a magnetic flux $\Phi$ to cross the 
circle. Then we identify $\theta = 2 \pi q \Phi$ (Aharonov-Bohm effect).
Unlike QCD, we can evaluate in the present toy model the effect
of $\theta >0$ precisely, see Section 4.\\

We now discretise the period $T$ in $L$ equal steps. We use
lattice units, {\it i.e.}\ we set the step length $T/L = 1$.
We denote $\varphi (t = j) = \varphi_{j}$, and
periodicity implies $\varphi_{j + L} \equiv \varphi_{j}$.
The standard action on this temporal lattice reads
\be
S_{s} [\varphi ] = m \sum_{j=1}^{L} \ [ 1 - \cos
( \varphi_{j+1} -\varphi_{j} ) ] \ .
\ee
In contrast, the perfect lattice action \cite{qrot}, which is obtained
from an infinite iteration of renormalisation group 
transformations,\footnote{For scalar particles in field theory
this method is discussed in Refs.\ \cite{scal}.}
distinguishes the topological sector that the particle may enter
at time slice $j$
\be
S_{p} [\varphi , \nu ] = \frac{m}{2} \sum_{j=1}^{L} \ [ \varphi_{j+1} 
- \varphi_{j} + 2 \pi \nu_{j}]^{2} \ .
\ee
In the functional integral all these sectors $\nu_{j} \in \Z$ are 
summed over, which reproduces the exact continuum result. \\

This discrete system has another interpretation as a {\em spin 
chain.} On each site $j$ a classical spin 
$\vec S_{j} = ( S^{(1)}_{j}, S^{(2)}_{j} ) $
of length $\vec S_{j}^{\, 2} = 1$ is attached.
This is the XY model, which has a global $O(2)$
symmetry. 
If we stay with periodic boundary conditions and
assume only nearest neighbour interactions
with some coupling $c$, we arrive at the partition function
\be
Z = {\rm Tr} \, e^{- \beta H [ \vec S ]} \ , \qquad
H  [ \vec S ] = - c \sum_{j=1}^{L} \vec S_{j+1} \cdot \vec S_{j} \ .
\ee
The trace means
the sum over all spin configurations
$ [ \vec S ] = (\vec S_{1}, \vec S_{2}, \dots , \vec S_{L})$, and
$\beta$ is an inverse temperature. If we identify the constants
as $\beta c = m$, we obtain the standard lattice path integral
of the quantum rotor at $\theta =0$ (up to an additive 
constant in the action); the angle $\varphi_{j}$ describes the direction
of the spin $\vec S_{j}$, and also the spin model can be generalised
by a $\theta$-term.\\

The standard discretised system does not have natural topological 
sectors, because all configurations can be continuously deformed into 
one another (in contrast to the case of continuous time). Still
one often introduces topologies, which is, however, ambiguous.
The most obvious option is the {\em geometric charge} \cite{geocharge},
which can be formulated analogously for instance in $N$-dimensional
$O(N+1)$ models, or in 4d Yang-Mills gauge theories
\cite{geochargeYM}. In our case, the geometric charge amounts to
\be  \label{geo}
\nu^{\rm (g)} = \frac{1}{2 \pi} \sum_{j=1}^{L} 
\Delta \varphi_{j}
\in \Z \ ,
\ee
where $\Delta \varphi_{j} = (\varphi_{j+1} -\varphi_{j}) 
\in (-\pi ,\pi ]$.

We will also consider alternative formulations.

\subsection{Observables}

We are going to extract the
correlation length $\xi$ as usual from the exponential decay of the
connected 2-point function (resp.\ a {\tt cosh} function due to
the periodic boundary conditions).
$\xi$ sets the scale of the system, and physically 
sensible results usually require
\be  \label{ineq}
1 \ll \xi \ll L \ . 
\ee
The first (second) inequality implies that discretisation artifacts
(finite size effects) are harmless.

Our observables are the topological and the magnetic susceptibility,
\bea
\chi_{t} &=& \frac{1}{L} \Big( \langle \nu^{2} \rangle -
\langle \nu \rangle^{2} \Big) \ ,  \label{chit} \\
\chi_{m} &=& \frac{1}{L} \Big( \langle \vec M^{2} \rangle -
\langle \vec M \rangle^{2} \Big) \ , \quad 
\vec M = \sum_{j=1}^{L} \vec S_{j} \label{chim} \ .
\eea

Let us assume the {\tt cosh} shape of the correlation
functions to hold at all distances. Then $\chi_{m}$ 
can be computed 
as follows,
\bea
\chi_{m} &=& \frac{1}{L} \Big\langle \ \Big( \sum_{j=1}^{L} 
\vec S_{j} \Big)^{2} \Big\rangle
= 1 +  \sum_{j = 2}^{L} \langle \vec S_{1} \vec S_{j} 
\rangle \nn \\
& \simeq & 1 + \sum_{j = 2}^{L} \Big[ e^{-(j-1)/\xi} + e^{-(L+1-j)/\xi} \Big]
= 
2 \ \frac{1 - e^{-L / \xi}}{1 - e^{-1/\xi}} - 1 \ .
\label{chimeq}
\eea
We now assume in addition the inequalities (\ref{ineq}) to hold.
In fact our simulations --- to be presented below ---
were performed consistently\footnote{We refer here to the correlation 
length $\xi$ at $\theta=0$.} at $L / \xi \approx 20$.
Thus the term $e^{-L / \xi}$ can be safely neglected, which leads to
\be  \label{chimapprox}
\frac{\chi_{m}}{\xi} \simeq 2 + \frac{1}{6 \xi^{2}} + O(\xi^{-3}) \ .
\ee

\section{A meron cluster simulation of the XY model}

\subsection{The algorithm}

We start by briefly reviewing the multi-cluster algorithm for $O(N)$
models \cite{Wolff,EdSo}, and in particular its extension to a meron 
cluster algorithm \cite{merons}. 

A step of the multi-cluster algorithm begins by building 
{\em clusters,} which are sets of neighbouring spins. 
To this end, a random direction $\vec{r}$ is chosen 
in an isotropic way ($| \vec r | =1$), 
and each spin $\vec{S}_{j}$ is split into $\vec{S}_{j}^{\parallel} = 
(\vec{S}_{j} \cdot \vec{r}) \, \vec{r}$ and $\vec{S}_{j}^{\perp} = 
\vec{S}_{j} - \vec{S}_{j}^{\parallel}$. A virtual bond is set between 
the sites $j$ and $j+1$ with the probability

\begin{equation}  \label{bonds}
p_{j} = \left\{\begin{array}{ll}
1 - e^{- 2 \beta \vec{S}_{j}^{\parallel}\vec{S}_{j+1}^{\parallel}} 
&\textnormal{\quad if \quad } \vec{S}_{j}^{\parallel} \cdot
\vec{S}_{j+1}^{\parallel} > 0 \\
0 &\textnormal{\quad otherwise \ .}
\end{array}
\right.
\end{equation}

Then a cluster is composed of neighbouring spins connected by bonds; 
it may also consist of a single spin, if the latter is disconnected. 
A step of the algorithm ends with ``flipping'' each cluster with 
the probability $1/2$ \cite{SW}. Flipping a cluster
means that all its spins are mirrored
at the plane perpendicular to~$\vec r$,
$\quad \vec{S}_{j}^{\parallel} \rightarrow -\vec{S}_{j}^{\parallel}, 
\quad \vec{S}_{j}^{\perp} \rightarrow \vec{S}_{j}^{\perp}$.
This algorithm respects ergodicity and detailed balance \cite{Wolff}.

Let us numerate the clusters with $c=1,2, \dots$.
A topological charge $Q_{c}$ can be assigned to each cluster 
based on the difference of the total topological charge $\nu$
(\textit{i.e.}\ the winding number) 
of the chain when the cluster is in its initial 
orientation, and after it has been flipped \cite{merons},
\be
Q_{c} = \frac{\nu_{c {\rm -initial}} - \nu_{c {\rm -flipped}}}{2} \ .
\ee
On the right-hand-side
we use the geometric charge (\ref{geo}).
$Q_{c}$ is a half-integer, which
remains unchanged if any other clusters 
are flipped, so it is determined locally \cite{Thomas}.
To illustrate this important property, 
let $\vec{S}_{j}$ \dots $\vec{S}_{k}$ be the spins of a specific cluster. 
We denote the sum of the relative 
angles $\Delta \varphi_{j} \in (- \pi , \pi ]$ (cf.\ eq.\ (\ref{geo}))
between the successive neighbouring spins $l$ to $n$ 
as $\widehat{\vec{S}_{l} \vec{S}_{n}}$, 
and a flipped spin is written as $\vec{S}'_{l}$. 
The cluster charge only depends on its 
boundary spins and the two neighbouring spins
of the adjacent clusters,
\begin{equation} \label{eq:Qc}
Q_{c} = 
\frac{1}{4\pi} \,
\bigg[ \, \widehat{\vec{S}_{j}\vec{S}_{k}} + \widehat{\vec{S}_{k}\vec{S}_{k+1}}
+ \widehat{\vec{S}_{j-1}\vec{S}_{j}} - \widehat{\vec{S}'_{j}\vec{S}'_{k}}
- \widehat{\vec{S}_{k}'\vec{S}_{k+1}} - \widehat{\vec{S}_{j-1}\vec{S}_{j}'} 
\, \bigg] \ .
\end{equation}
The charge is the same, regardless whether the neighbouring
cluster is flipped or not, provided that 
$\widehat{\vec{S}_{j-1}\vec{S}_{j}} - \widehat{\vec{S}_{j-1}\vec{S}'_{j}} 
= \widehat{\vec{S}'_{j-1}\vec{S}_{j}} - 
\widehat{\vec{S}'_{j-1}\vec{S}'_{j}}$.
In fact, this holds generally, which is easy to show
by distinguishing different cases of the angles 
$\widehat{\vec{S}_{j}\vec{S}_{j-1}}$ and 
$\widehat{\vec{S}_{j}\vec{S}'_{j-1}}$ . 

It is a peculiarity of the spin chain that
there cannot be any loop inside a cluster, as we see from prescription
(\ref{bonds}). Thus the cluster charges are limited to the values 
$1/2$, $0$ and $-1/2$, and the corresponding clusters are denoted as 
{\em meron,} neutral cluster and {\em anti-meron,}
respectively.

The property that $Q_{c}$ is determined locally for each 
cluster\footnote{In higher dimensions this vital property can only 
be achieved by imposing constraints on the maximal angles between
neighbouring spins \cite{merons}, cf.\ Section 1.}
enables us to construct an \textit{improved estimator,}
which will be applied as a powerful tool in this work. 
With $N_c$ clusters, $2^{N_c}$ configurations can be 
obtained by cluster flips, which could enter the statistics
(without the need for a Metropolis accept-reject step). 
In practice it is not optimal --- or not even possible ---
to include all of them 
(we encountered $N_{c}$ values up to $40$), unless 
this average can be evaluated analytically. Of course these
configurations are
not fully independent because they are all affiliated to the same
direction $\vec r$. 

\subsection{Cluster statistics}

It is common lore that the ``characteristic''
cluster size follows the correlation length. Taking a close
look at this property, we found that the statistical distribution
of cluster with length $s$ can be fitted well to a sum of
three exponentials, $\sum_{i=1}^{3} c_{i} \exp (- s / s_{i})$.
For large clusters the first term is dominant.
Its decay is given by $s_{1} = 1.000(1) \xi$, 
in precise agreement with the expectation.
This supports the interpretation of the clusters as physical
degrees of freedom, which is the basis of the meron picture
employed here.
\begin{figure}[h!]
 \centering
  \includegraphics[angle=0,width=.5\linewidth]{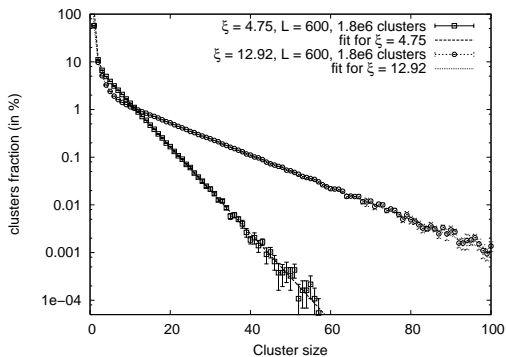}
 \caption{{\it The statistical distribution of the cluster sizes, measured
at two different correlation lengths. The density of large clusters
decays exponentially, where the characteristic size coincides with
the correlation length.}}
\label{clustersizefig}
\end{figure}
At smaller cluster sizes the curve is steeper than the
first exponential alone, see Figure \ref{clustersizefig}.
The leading sub-dominant exponential has a short range of
$s_{2} = 0.111(2) \xi$. We add that the continuum limit 
leads to a stable fraction of $58.0(1) \%$
clusters of the minimal size $s=1$.\\

Next we consider the fraction of merons among
the clusters. At large $\xi$ it amounts to
$0.1581(5) \, \xi^{-0.542(2)}$ (of course the same holds for the
anti-merons). Obviously large clusters have a
higher probability to carry topological charge. In the limit
of a very large size $s$ one finds $25 \%$ merons; around
$s = 2 \, \xi$ one already arrives approximately at 
this asymptotic number.

To provide an intuitive argument for this property,
let us assume for instance a direction $\vec r = (0,1)$,
and we measure the spin angles $\varphi_{j} \in (- \pi , \pi]$ relative
to the $x$-axis. Again we consider some cluster with the spins
$\vec S_{l} \dots \vec S_{k}$ and we assume $\varphi_{l} \in (0,\pi /2)$.
Now the spin angles of this cluster describe a discrete path 
in $(0,\pi )$. For large $\xi$ (resp.\ large $\beta$)
the relative angles $\Delta \varphi_{j}$ of adjacent spins
are small. In particular we can assume the continuum limit
to be approached to a point where the probability for any
$| \Delta \varphi_{j}| \geq \pi /2$ is negligible.
Hence for small clusters $\varphi_{k}$ is likely to be
in $(0, \pi /2)$ as well, so that the cluster is neutral. However,
in very large clusters $\varphi_{l}$ becomes irrelevant for the
endpoint $\varphi_{k}$, which can be in $(0 , \pi / 2)$ or
in $(\pi /2 , \pi)$ with equal probability.
This implies an equal number of neutral clusters and merons. 
Analogously, large clusters with 
$\varphi_{l} \in (\pi /2 , \pi )$ are equally likely to be neutral
or anti-merons.

To quantify the increase of the meron density as $s$ rises,
we specified three densities and Table \ref{tabclust}
displays the corresponding sizes $s$.

\begin{table}
\begin{center}
\begin{tabular}{|c|c|}
\hline
meron density & cluster size  $s$ \\
\hline
\hline
2.5 \%  & $0.401(5) \, \xi - 0.10(3)$ \\
\hline
12.5 \% & $0.644(3) \, \xi + 0.23(3)$ \\
\hline
22.5 \% & $1.16(1) \, \xi + 0.93(7)$ \\
\hline
\end{tabular}
\end{center}
\caption{{\it The cluster sizes corresponding to three 
specific meron densities.}}
\label{tabclust}
\end{table}

\subsection{Efficiency}

For comparison, we consider as a unit of computation
time a process that could 
modify the whole chain: for the multi-cluster algorithm, this is
what we have described before as one algorithmic step; 
for Metropolis, it means one sweep to tackle 
each spin in the chain. We repeat that 
we performed our tests at a chain length $L \simeq 20 \cdot \xi$,
so that finite size effects are strongly suppressed.\footnote{Efficiency
studies in the two dimensional XY model, with multi-cluster and single 
cluster algorithms, were presented in Refs.\ \cite{EdSo,single}.}

\begin{figure}[h!]
 \centering
  \includegraphics[angle=0,width=.5\linewidth]{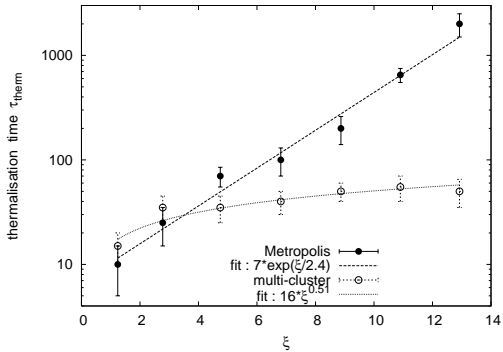}
 \caption{{\it The thermalisation time $\tau_{\rm therm}$ for the energy 
as a function of the correlation length $\xi$. In the multi-cluster
simulation $\tau_{\rm therm}$ increases only slowly
in $\xi$, which is in contrast to the Metropolis algorithm. 
The results are averaged over a variety of
cold and hot starts.}}
\label{thermalisation}
\end{figure}
First we consider the thermalisation time with respect to the energy.
For the Metropolis algorithm $\tau_{\rm therm}$ grows
exponentially with the correlation length, 
$\tau_{\rm therm} \simeq 7(2) \exp(\xi /2.4(2))$. 
On the other hand
the data for the multi-cluster algorithm follow
a power law, $\tau_{\rm therm} \simeq 16(2) \xi^{0.51(7)}$
as shown in Figure \ref{thermalisation}.
In particular this ensures a striking advantage at large $\xi$,
when we approach the continuum limit. 

\begin{figure}[h!]
\centering
\includegraphics[angle=0,width=.5\linewidth]{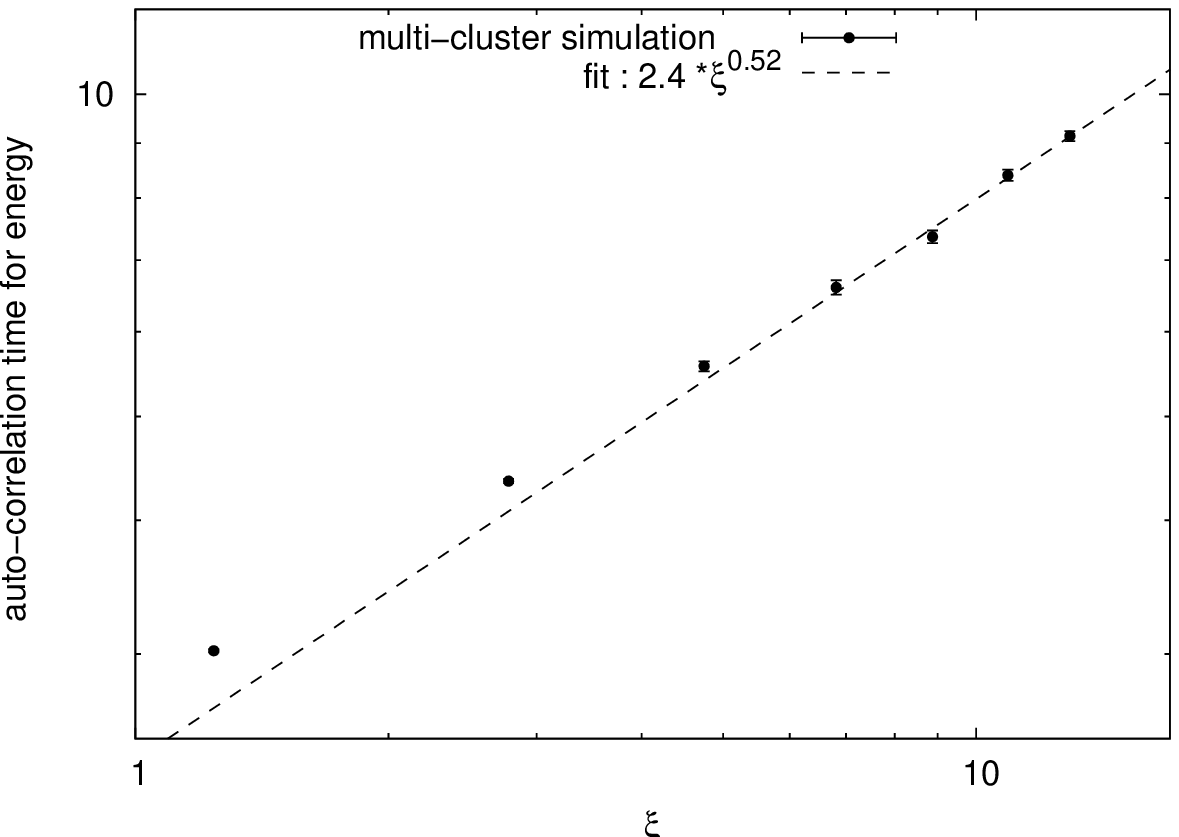}
\hspace*{-1mm}
\includegraphics[angle=0,width=.48\linewidth]{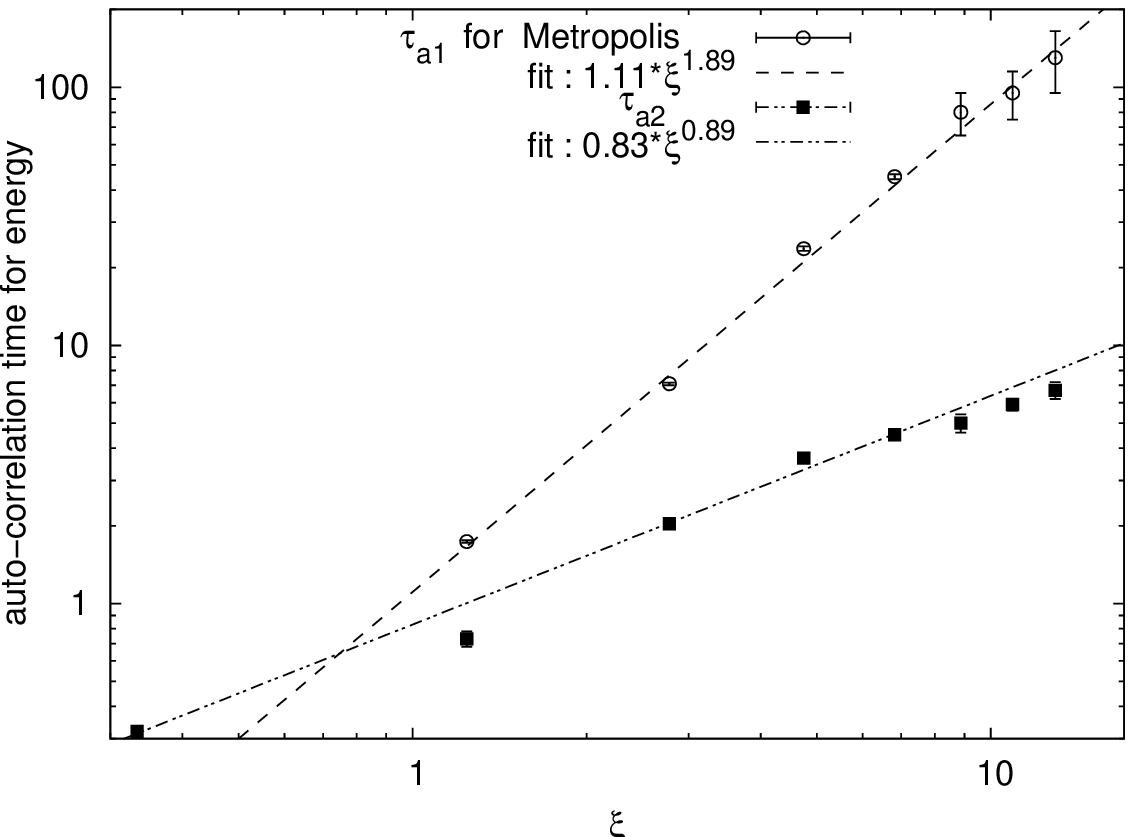}
\caption{{\it The auto-correlation time $\tau_{a}$ for the energy.
The plot on the left shows the result obtained with the multi-cluster 
algorithm, where we observe a modest increases $\tau_{a} \propto \xi ^{0.52}$.
On the right we show the Metropolis result.
It can be fitted with two exponentials, with a dominant exponent of $1.89$.}}
\label{MC_correl_energy}
\end{figure}

We proceed to the stage where the thermalisation is completed
and we consider now the (exponential) auto-correlation time $\tau_{a}$, 
again with respect to the energy. Hence
the auto-correlation function is fitted 
with $\exp (-\tau /\tau_{a} )$, where $\tau$ 
is the algorithmic time. The values of $\tau_{a}$ are 
plotted in Figure \ref{MC_correl_energy} (on the left)
at various~$\xi$ for the multi-cluster algorithm; 
we observe $\tau_{a} \propto \xi^{\gamma}$ 
with a dynamical critical exponent $\gamma = 0.52(3)$.

For the Metropolis simulation, the data can be fitted well with a 
sum of two exponentials, $ c \exp(- \tau / \tau_{a1}) + 
(1-c) \exp(- \tau / \tau_{a2} )$, 
and Figure \ref{MC_correl_energy} (on the right) shows the corresponding
results $\tau_{a1}$ and $\tau_{a2}$. The corresponding
critical exponents amount to
$1.89(6)$ and $0.89(4)$. What ultimately matters is
the dominant exponent $\approx 2$, which is reminiscent
of the random walk diffusion of local changes on the chain.

\begin{figure}[h!]
  \centering
\includegraphics[angle=0,width=.5\linewidth]{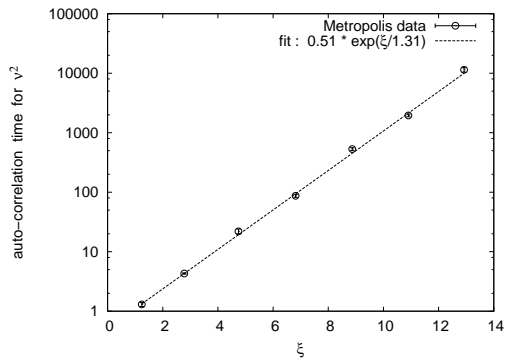}
\caption{{\it The auto-correlation time $\tau_{a}$ with respect to
the topological charge squared. The plot shows its exponential 
increase with $\xi$ for the Metropolis algorithm. In contrast,
it practically vanishes with the multi-cluster algorithm.}}
\label{Metro_correl_khi}
\end{figure}

For the (squared) topological charge $(\nu^{\rm (g)})^{ \, 2}$
(which is relevant for $\chi_{t}$), the growth of the
auto-correlation time is exponential with Metropolis, 
as Figure \ref{Metro_correl_khi} shows. On the other
hand, the auto-correlation 
practically vanishes with the multi-cluster algorithm. This de-correlation 
is due to the large-scale changes performed on the chain. The 
multi-cluster algorithm reveals here most clearly
its potential in overcoming the critical slowing down.

\section{Results for the observables}

\subsection{Correlation length $\xi$}

\begin{figure}[h!]
  \centering
\includegraphics[angle=0,width=.5\linewidth]{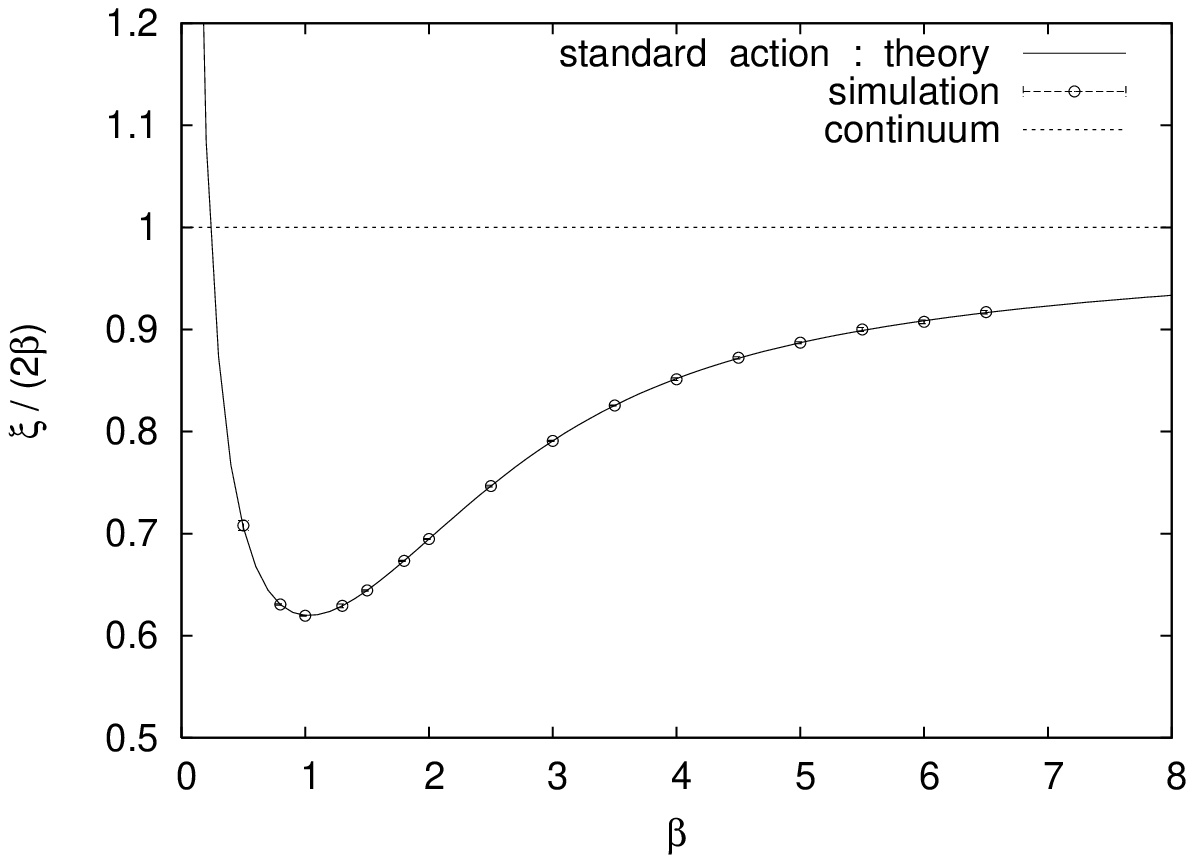}
\hspace*{-3mm}
\includegraphics[angle=0,width=.5\linewidth]{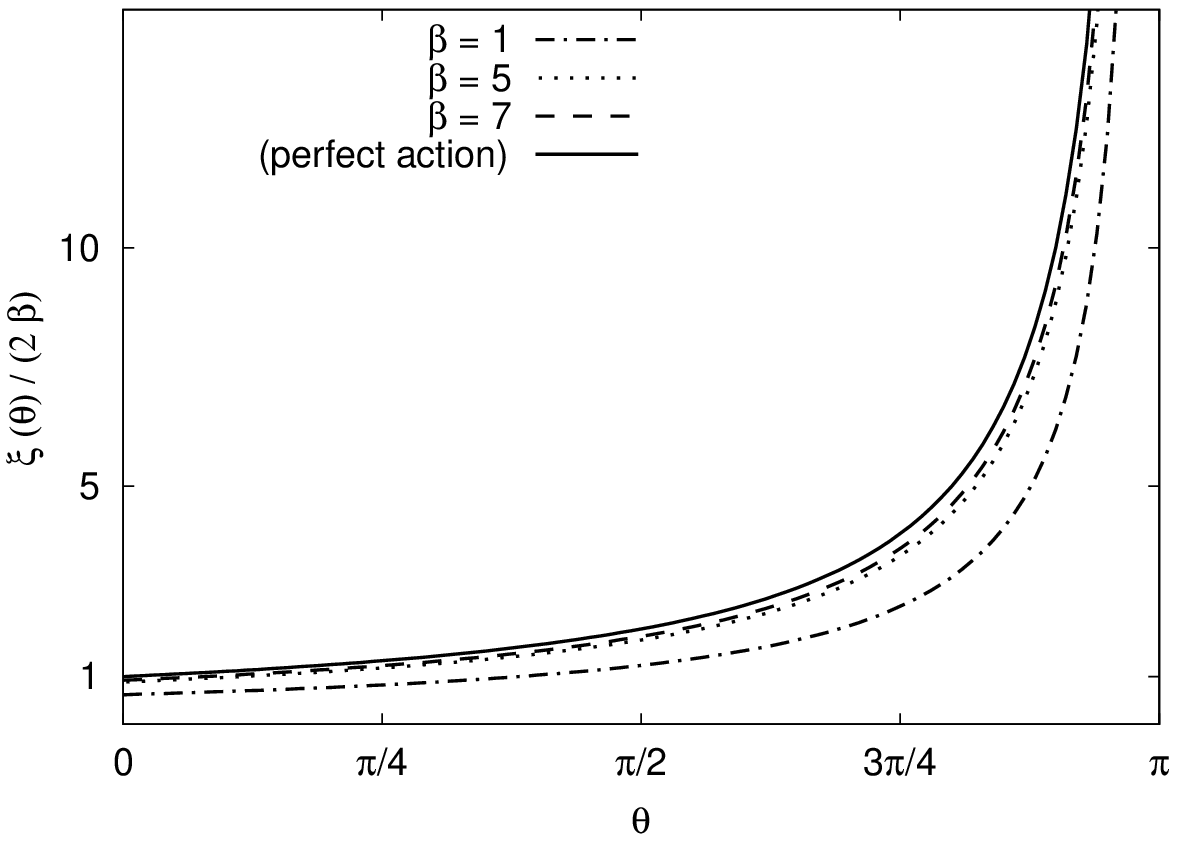}
 \caption{{\it The correlation length as a function of $\beta$ for
the standard action.
The simulation results at $\theta=0$ (on the left) are in 
excellent agreement with the theoretical prediction \cite{qrot}.
At finite $\theta$ (shown on the right) and increasing $\beta$
they converge towards the perfect action
result, which describes the system in the continuum.}}
\label{xi_beta}
\end{figure}

In Subsection 3.2 the correlation length (at $\theta = 0$)
has been anticipated.
We now consider its relation to the inverse temperature $\beta$.
The numerical and theoretical \cite{qrot} results
with the standard action match perfectly, as 
the plot in Figure \ref{xi_beta} on the left shows.
On the right we add results for $\xi$
at non-zero vacuum angles $\theta$. For increasing $\beta$
the correlation length of the standard action approaches
the perfect action value, $\xi / (2 \beta ) = 1/ (1 - \theta /\pi)$.
A divergence at $\theta \to \pi$ has also been observed in 
the 2d $O(3)$ model \cite{merons}.

\subsection{Topological susceptibility $\chi_t$}

\begin{figure}[h!]
  \centering
  \includegraphics[angle=0,width=.5\linewidth]{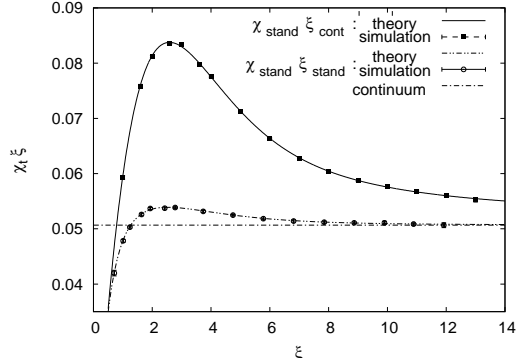}
 \caption{{\it The numerical results for the topological
susceptibility $\chi_t$ (at $\, \theta =0$), in precise agreement 
with the theoretical prediction for the standard lattice
action \cite{qrot}. The result for the perfect action coincides 
with the continuum susceptibility at any correlation length.}}
\label{khi_xsi}
\end{figure}

Figure \ref{khi_xsi} shows the accurate agreement of the measured
topological susceptibility (eq.\ (\ref{chit})) at $\theta =0$
with the theoretical formula of Ref.\ \cite{qrot}.

\begin{figure}[h!]
\centering
 \includegraphics[angle=0,width=.5\linewidth]{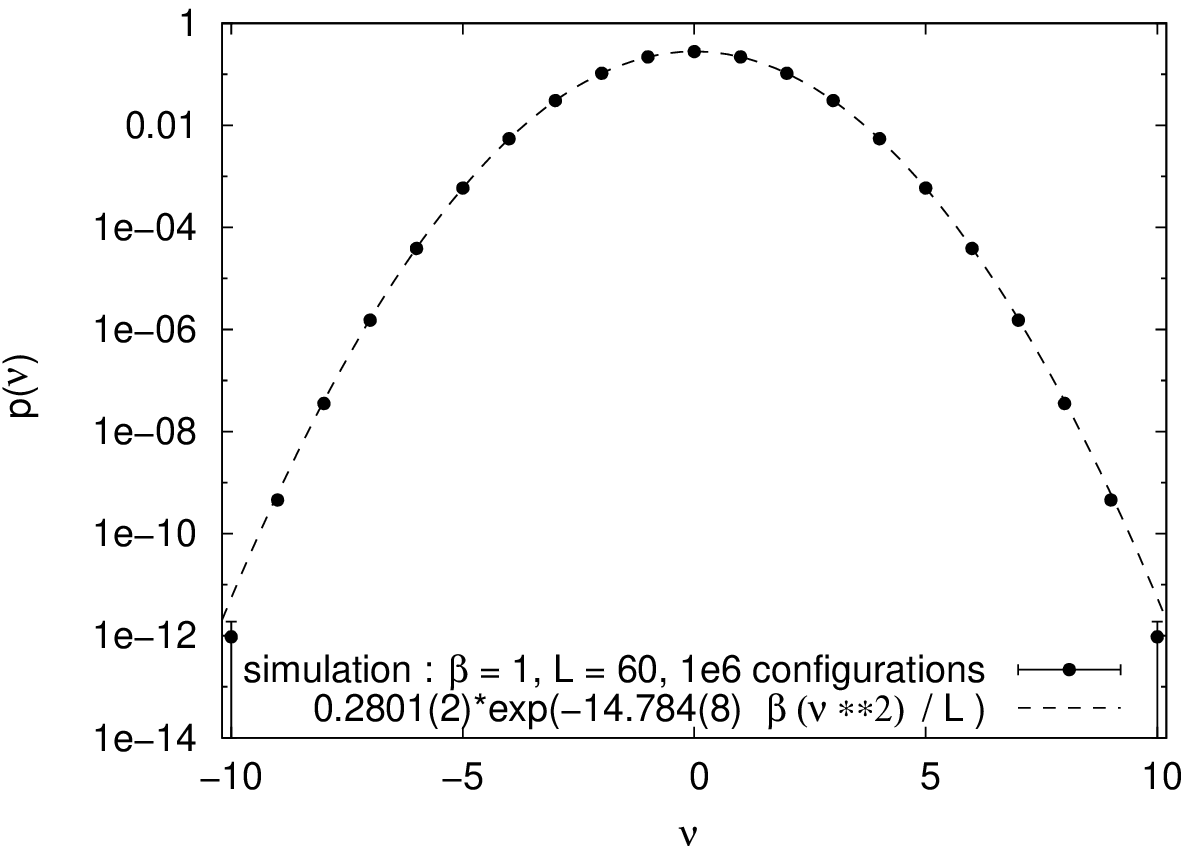}
\hspace*{-3mm}
 \includegraphics[angle=0,width=.5\linewidth]{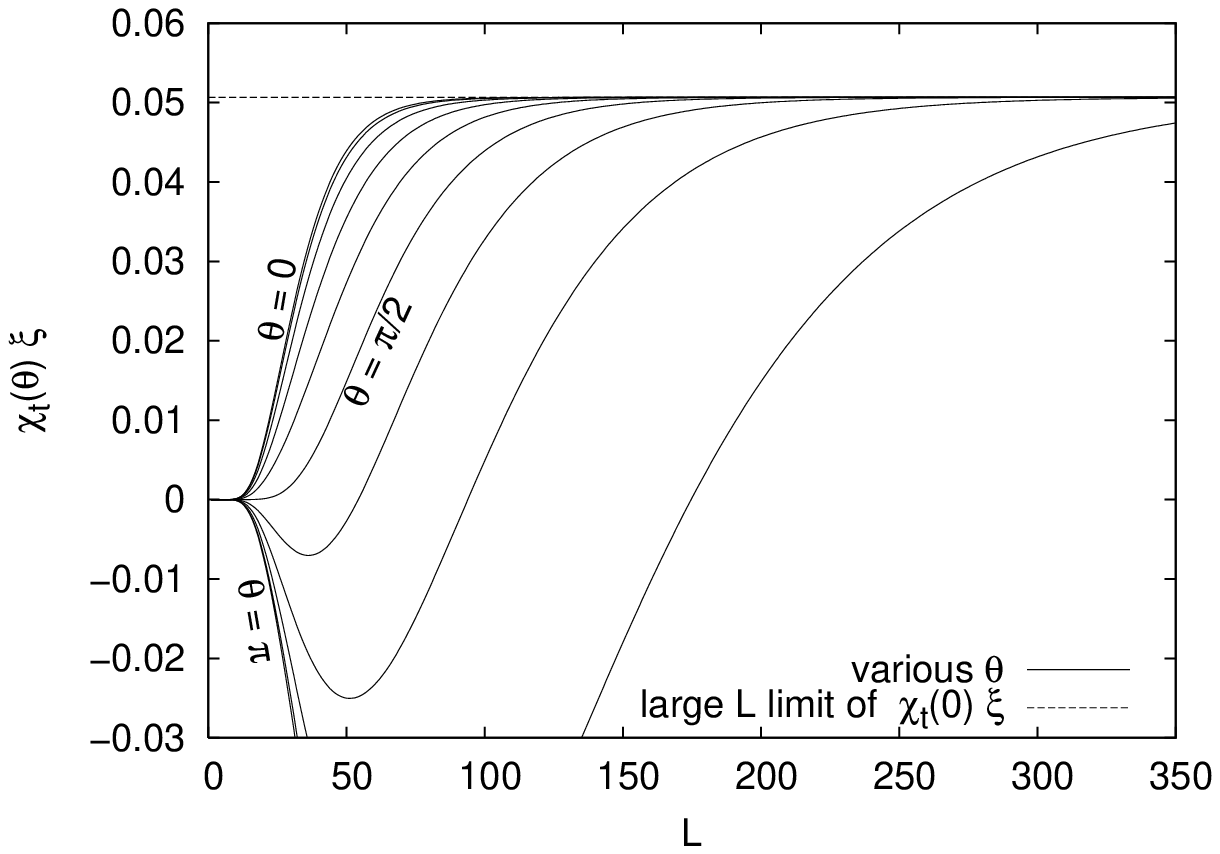}
\caption{{\it On the left: the probability distribution 
for the topological charges, $p(\nu )$, which follows
a Gaussian. An improved estimator is mandatory to capture
even tiny probabilities. 
On the right: $\chi_t(\theta) \xi $ at different vacuum angles, 
against the length $L$ of the spin chain. At large $L$ it becomes 
independent of $\theta$ (except for $\theta = \pi$).}}
\label{P_of_Q}
\end{figure}

To calculate $\chi_t (\theta )$ also at $\theta \neq 0$, the probability 
$p(\nu )$ is needed to an extremely high precision. It proves to be 
Gaussian, see Figure \ref{P_of_Q} (on the left).
An improved estimator can be used here which captures all cluster 
orientations by simple combinatorics. 
This enhances the statistics drastically, and
it allows us to reach probabilities of $O(10^{-12})$ with only 
one million really generated configurations.
Subsequently we can use an analytic expression for
the distribution $p(\nu )$.

The results for the $\theta$-dependent susceptibility
$\chi_t(\theta) = \frac{1}{L} \left(\langle \nu^{2} \rangle_{\theta} 
- \langle \nu \rangle^2_{\theta} \right)$
were obtained by relying on the Gaussian distribution
$p(\nu )$ that we identified. Note that both terms contribute.
$\chi_t(\theta)$ is real due to the parity symmetry, which implies
the symmetry in the sign of $\nu$.
Figure \ref{P_of_Q} (on the right) shows its dependence on $L$ for
various values of $\theta$. 
(To set the scale, we still refer to $\xi$ at $\theta =0$.)
It converges for large $L$
to the value of $\chi_{t}(0)$, for any $\theta \neq \pi$. 
This convergence slows down as $\theta$ rises, 
and it collapses at $\theta = \pi$.

Such precise results for a system with a complex action
are very rare; other examples with a $\theta$-term were obtained
with the meron cluster algorithm for the 2d $O(3)$ model
\cite{merons} and for an $SU(N)$ quantum spin ladder 
\cite{BPRW}.\footnote{For alternative methods to simulate
models with a $\theta$-term,
see {\it e.g.}\ Refs.\ \cite{numtheta}.}

\subsection{The topological susceptibility from cooling}

\begin{figure}[h!]
  \centering
\includegraphics[angle=0,width=.5\linewidth]{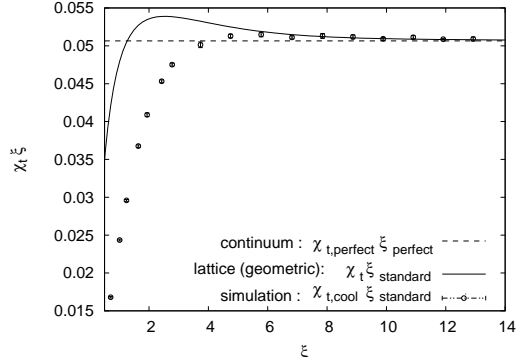}
 \caption{{\it The topological susceptibility obtained from ``cooling''.
For increasing $\xi$ it converges to the continuum value with the same
speed as the use of the geometrical charge on the uncooled configurations.}}
\label{khi_cool}
\end{figure}

For comparison we also consider this susceptibility
based on topological charges obtained from ``cooling'' \cite{cool}; 
we denote it as $\chi_{t,{\rm cool}}$. To this end, the chain 
is smoothed before a measurement: a spin is chosen at random and 
rotated so that the action is minimised. This process is iterated 
until it converges. In lattice gauge theory a long cooling process
(for the plaquettes) ultimately leads to the trivial configuration,
so one tries to read off a topological charge from some
earlier plateau (for instance by monitoring the energy).
Here the situation is simpler because the cooled configuration
stabilises, so $\nu^{\rm (cool)}$ is (in this sense)
unambiguous. The question remains how it is
related to the original configuration, which has the correct
statistical weight (unlike the cooled configuration).
We may compare the cooled charge to the original
geometrical charge $\nu^{\rm (g)}$, with the obvious inequality 
$|\nu^{\rm (cool)}| \leq |\nu^{\rm (g)}|$. The ultimate 
criterion is how well $\chi_{t,{\rm cool}}$ approximates the
continuum value of $\chi_{t}$. The result is plotted Figure
\ref{khi_cool}. The convergence of $\chi_{t,{\rm cool}}$ 
to the continuum limit is just as quick as
for the geometrical charge without cooling.

\subsection{Magnetic susceptibility $\chi_m$}

\begin{figure}[h!]
\centering
\includegraphics[angle=0,width=.5\linewidth]{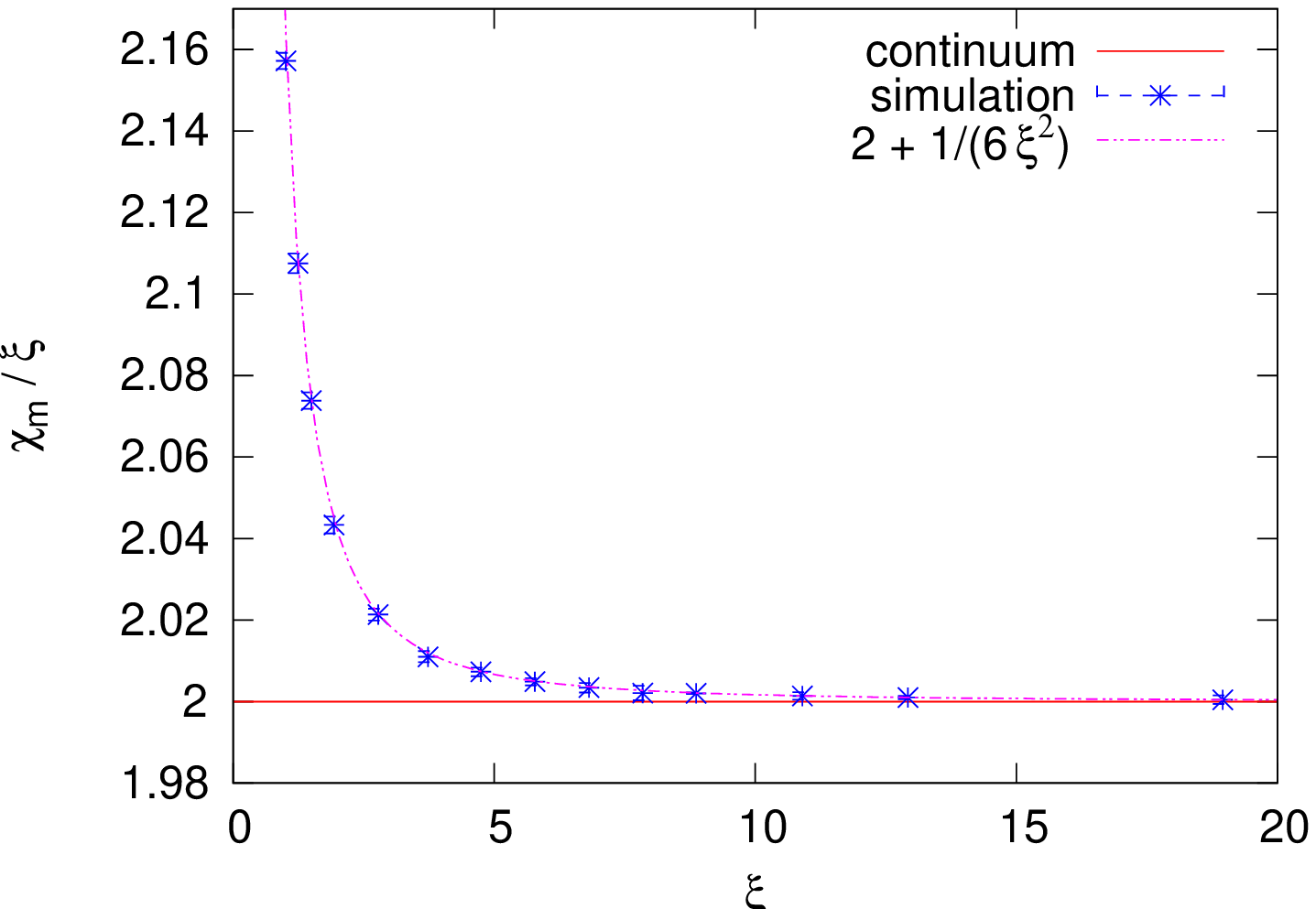}
\hspace*{-3mm}
\includegraphics[angle=0,width=.5\linewidth]{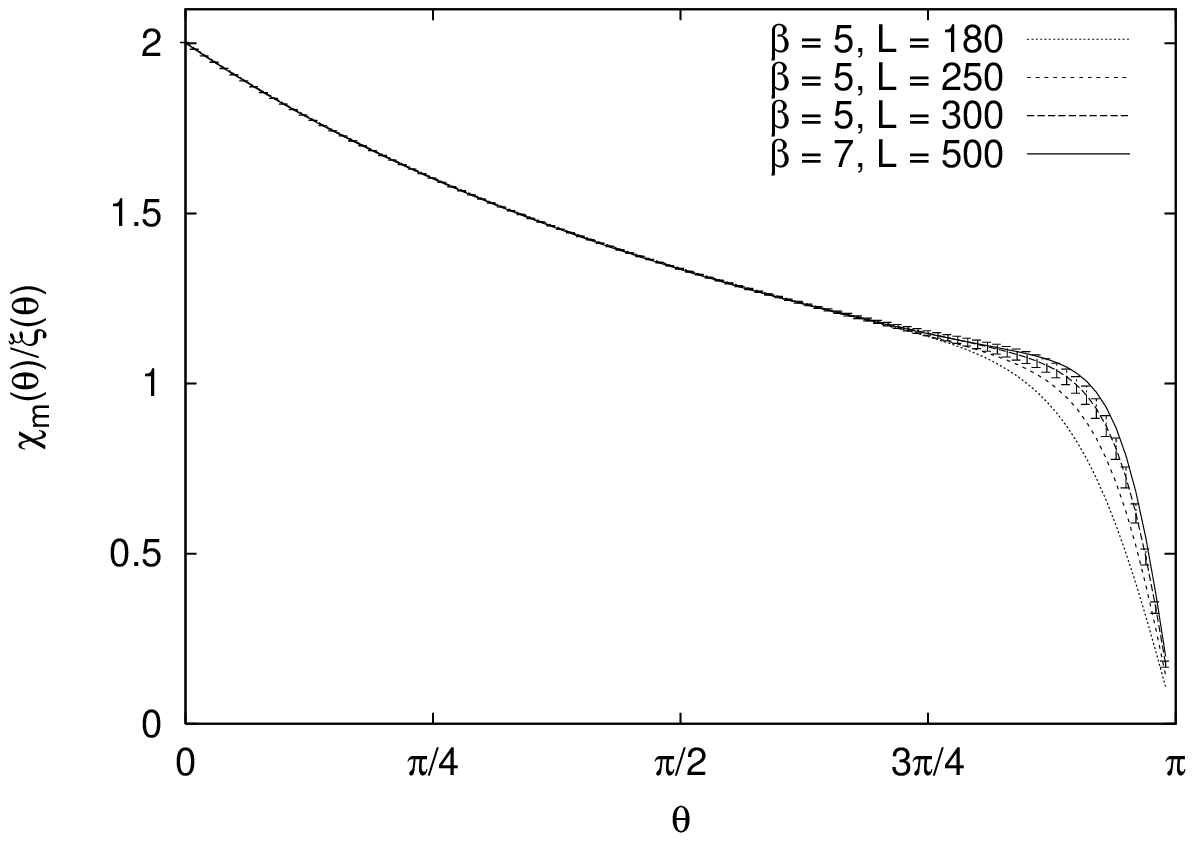}
\caption{{\it The numerical results for the magnetic
susceptibility. On the left we show the result at $\theta =0$,
which follows the formula (\ref{chimapprox}) 
to an excellent accuracy. The plot on the right
is our result for $\chi_m (\theta) / \xi (\theta )$. 
The data are reliable over the plotted range, before getting too
close to $\pi$. This result was accessible
with the multi-cluster algorithm along with an improved estimator.}}
\label{chi_m}
\end{figure}

Figure \ref{chi_m} (on the left) shows our results for the 
magnetic susceptibility $\chi_{m}$, see eq.\ (\ref{chim}). They 
are in excellent agreement with the approximation given in
eq.\ (\ref{chimapprox}). We mention that $\chi_{m}$ corresponds 
to the mean cluster size when one employs the single cluster 
algorithm \cite{Wolff}, which favours larger clusters than the
multi-cluster method.

At non-zero $\theta$, the result emerges from re-weighting
the data of $\theta =0$ according to the prescription in
Ref.\ \cite{thetasimu}. 
The results are shown in Figure \ref{chi_m} on the right.
We applied an improved estimator
by flipping a number of clusters.
Here we do consider the $\theta$-dependence of the
correlation length, which is used to build the dimensionless
ratio plotted in Figure \ref{chi_m}. It
grows rapidly as $\theta$ approaches $\pi$,
cf.\ Figure \ref{xi_beta}, hence we involved quite long spin chains.
Once $\xi (\theta )$ exceeds $L$, the considered ratio drops towards $0$,
as we see from eq.\ (\ref{chimeq}). However, for a continuum
limit $\theta \to \pi$ which respects $L \gg \xi$ the plateau at
relatively large $\theta$
suggests a value $\chi_{m}(\theta) / \xi (\theta) \approx 1$.

In one case, $\beta=5$, $L=300$, we indicate jackknife errors. 
In the other cases they are similar, {\it i.e.}\ again very small
up to the close vicinity of $\theta = \pi$. This is very remarkable in
view of the generic difficulties to obtain numerical results for models
with a significant imaginary part in the Euclidean action.


The statistics includes several millions of 
configurations, and it is still enhanced 
thanks to the improved estimator.
In particular, statistics in a fixed topological sector
can be cumulated by flipping neutral clusters.
In this way it could be simulated within a few weeks on a
2~GHz machine; with the Metropolis algorithm this measurement
is hardly feasible.

\section{Conclusions and outlook}

We presented a new and successful application of
the meron cluster algorithm. In the
spin chains that we considered, it provides 
precise simulation results in a highly efficient way.
We assigned a half-integer topological charge to each cluster, 
which is the basis of a powerful improved estimator.
This yields accurate result for the present toy model
with a $\theta$-term --- an issue, which is still outstanding
in QCD.

Here it was possible to suppress the notorious
problem of critical slowing down. With respect to the
topological charge we overcome this problem completely.
Regarding the energy, the dynamical critical exponent is
reduced by almost a factor of $4$. \\

This provides a strong motivation to search for applications of this 
technique also in higher dimensions, {\it i.e.}\ in field theoretic 
models. In particular an application of the meron cluster
algorithm in the 3d $O(4)$ model appears promising --- that
model can be interpreted as an effective description of QCD with
two light quark flavours at high temperature. 

We add a very rough estimate about the feasibility of that project. 
If $\xi (\theta )$ behaves similarly to Figure \ref{xi_beta},
we control the finite size effects quite well up to $\theta \approx
0.9 \, \pi$ for instance with $L =32$. 
Compared to the spin chains that we considered to measure
$\chi_m(\theta ) / \xi (\theta)$, this implies
a factor of $O(100)$ for the number of lattice sites (along with a 
factor of $6$ for the generators of the symmetry group). 
Tiny error bars as we obtained in Figure \ref{chi_m} (on the right) may be 
relaxed without problems, say by a factor $\gsim 3$, so that
the required statistics decreases by an order of magnitude.
Comparing now to the computational effort which was necessary in $d=1$
(cf.\ last paragraph in Section 4), and considering the option to use
a number of processors simultaneously, we estimate that the 3d $O(4)$ model
at finite $\theta$ can be solved to a good precision
with the meron cluster algorithm within less than one year.\\

\noindent
{\bf Acknowledgements:} {\em We thank Michael M\"{u}ller-Preu\ss ker,
Andr\'{e} Sternbeck, Jan Volkholz, Uwe-Jens Wiese and Ulli Wolff for 
useful comments. The computations were performed on a
PC cluster at the Humboldt-Universi\-t\"{a}t zu Berlin.}

\end{document}